\documentclass{PoS}

\usepackage{amsmath}
\usepackage{adjustbox}
\usepackage{booktabs}
\usepackage{sidecap}
\usepackage{floatrow}
\usepackage{url}
\newfloatcommand{capbtabbox}{table}[][\FBwidth]

\newcommand{\fig}{Fig.}

\newcommand{\sibyll}[1]{{\sc Sibyll#1}}

\newcommand{\dpmjet}[1]{{\sc DPMJET#1}}

\newcommand{\mceq}{{\sc MCEq}}

\newcommand{\figu}[1]{\fig~\ref{fig:#1}}

\newcommand{\bi}{\begin{itemize}}
\newcommand{\ei}{\end{itemize}}

\title{Calibration of atmospheric neutrino flux calculations using cosmic muon flux and charge ratio measurements}

\ShortTitle{Combined atmospheric muon fit}

\author{\speaker{Juan-Pablo Y\'{a}\~{n}ez}$^\dagger$, Anatoli Fedynitch$^\star$ and Tyler Montgomery\\
		Dept.\ of Physics, University of Alberta, Edmonton, Alberta, Canada T6G 2E1\\
        E-mail:  \email{j.p.yanez@ualberta.ca}$^\dagger$, \email{anatoli.fedynitch@ualberta.ca}$^\star$}

\abstract{The general features of the neutrino flux from cosmic ray interactions in the Earth's atmosphere are well characterized. However, the absolute precision of calculations is still insufficient and the uncertainty from the modeling of hadronic interactions in the very forward region remains a major limitation. In this work, we benchmark the current generation hadronic models using high-precision atmospheric muon calculations from a few GeV to multiple TeV energies provided by the MCEq code. We derive corrections to hadronic models using publicly available measurements of the flux and charge ratio of atmospheric muons from surface and underground detectors. When combining data, the experimental uncertainties are taken into account. We discuss the calibration method and the strength of the derived constraints.}

\FullConference{36th International Cosmic Ray Conference -ICRC2019-\\
		July 24th - August 1st, 2019\\
		Madison, WI, U.S.A.}
		
\begin{document}

\section{Introduction}

Atmospheric neutrinos have contributed significantly to our understanding of neutrino properties and, in the emerging field of neutrino astronomy, they constitute the main source of background. A high precision calculation of their flux is, therefore, paramount to achieve the ambitious physics goals of future large-volume neutrino detectors, such as Hyper-Kamiokande and IceCube-Gen2.

Hadronic interactions play a crucial role in flux calculations. Of particular importance is the modeling of light meson production in the very forward phase-space, {\it i.e.}~of secondary particles produced at very small scattering angles that would usually escape the detection at modern colliders due to the presence of a beam pipe. Fixed target experiments can shed light on some specific regions of this phase-space, but they can only be performed for a limited number of targets and energies.

Theoretically, this phase-space is the domain of non-perturbative QCD implying the absence of robust computational methods that would allow one to predict the required particle spectra from first principles. There is, however, a number of phenomenological models implemented as Monte Carlo event generators in use, and in the past decades they have evolved into sophisticated but still imperfect tools. The different ideologies of these models, or of purely data-driven methods \cite{Engel:2001qi}, constitute to a large extent the main source uncertainty in atmospheric lepton flux calculations \cite{Barr:2006it,Fedynitch:2018vfe}, and cannot be easily reduced due to the absence of more constraining data.

One possible way to reduce their impact on neutrino flux calculations is to exploit the correlation with the flux of atmospheric muons that originate in the same decays of charged pions and kaons as the neutrinos. In comparison to neutrinos, the flux of $\mu^+$ and $\mu^-$ can be measured relatively precisely with small spectrometers and therefore it can serve as a calibration source for the models involved in the flux computations. This method \cite{Honda:2006qj} has been employed in the most successful neutrino flux calculation by Honda et al.~\cite{Honda:2015fha}.

A large number of experiments have performed measurements of both the flux and the charge ratio $(\mu^+/\mu^-)$. The measurements are affected by specific experimental conditions that can range from atmospheric effects to magnet alignment for charge identification, which makes their comparison complicated. Here, for the first time, we present a comprehensive analysis of publicly available atmospheric muon data by combining several experiments with a detailed treatment of their systematic uncertainties. Our aim is to derive corrections for hadronic interaction models and test their impact on the resulting {\it calibrated} neutrino fluxes.

\section{Flux model}
\label{sec:flux_computation}
We employ the cascade code \mceq{} \cite{Fedynitch:2015zma}\footnote{https://github.com/afedynitch/MCEq} to perform the calculations of the lepton fluxes. Despite a high computational speed, the code is not fast enough to be directly involved in a minimization. Therefore, we pre-compute a database of fluxes individually taking into account the conditions of each measurement. These conditions include the reported zenith angles, the altitude, the atmosphere (averaging over the duration of data-taking) using NRLMSISE-00 \cite{Picone:2002go}, and the variable in which the spectrum is reported (momentum or energy).

\mceq{} contains functions that modify the multiplicity distributions of pions and kaons for proton-air and neutron-air interactions, starting from a baseline interaction model. Currently, we use \sibyll{-2.3c} \cite{Riehn:2017mfm} and \dpmjet{-III-19.1} \cite{Fedynitch:2015kcn} in combination with GSF as the model for the cosmic ray flux at the top of the atmosphere \cite{Dembinski:2017zsh}. GSF is a fit performed on a large set of cosmic ray data relying on a minimal set of assumptions and comes with error estimates. These have not yet been incorporated into this work and are the subject of a future study.

To parameterize the impact of hadronic uncertainties, we subdivide the particle production phase-space into discrete regions in projectile energy and secondary energy fraction exactly following Barr et al.~(see Fig.~3 in \cite{Barr:2006it}). While the full scheme contains $11 \times 2$ parameters $\mathcal{B}_i$ for pions and kaons of both charges in proton-air interactions, the contributing number of parameters at energies $>20$ GeV is eight (G, H, W and Y for each charge sign).
To propagate the effect of the modified particle yields, we use an original scheme that developed for the fast propagation of model errors impacting the lepton flux calculations \cite{Fedynitch:2018vfe}. In this scheme a Jacobian ``matrix'' is constructed by computing the first term of a Taylor expansion of the muon flux $\Phi(E_\mu)$ to a perturbation $\delta$ with respect to the variation of a single parameter $\mathcal{B}_i$:
\begin{equation}
    \frac{\partial \Phi(E_\mu)}{\partial \mathcal{B}_i} = \frac{\Phi(E_\mu,\mathcal{B}_i=1 + \delta) - \Phi(E_\mu,\mathcal{B}_i= 1 - \delta)}{2 \delta}.
\end{equation}
These Jacobians are interpolated and stored together with the unperturbed flux. The flux for each experimental site with the corrections, $\mathcal{B}_i$, applied can be computed easily from
\begin{equation}
    \Phi(E_\mu, \mathcal{B}_a, \mathcal{B}_b, \dots) = \Phi(E_\mu) + \sum_i \mathcal{B}_i  \frac{\partial \Phi(E_\mu)}{\partial \mathcal{B}_i}.
\end{equation}
Since the coupled cascade equations (see for instance \cite{Gaisser:2016uoy}) solved by \mceq{} are linear, this approach is exact and does not require higher order terms. The resulting database of fluxes and Jacobians for each experimental site is fast to evaluate for arbitrary combinations of $\mathcal{B}_i$ and can be directly used by minimizers. 

\section{Muon flux and charge ratio data}

We conducted a comprehensive literature survey to identify suitable measurements of muon fluxes and charge ratios. In order for a measurement to be included, it must have been published in a peer-reviewed journal with a detailed description of the measurement conditions and the systematic uncertainties. An incomplete description of the systematic uncertainties was the most frequent reason for discarding a measurement. Table \ref{table:experiments} shows the list of experiments that were selected for the analysis.

\begin{table}[!t]
\centering
\begin{adjustbox}{width=1\textwidth}
%\small
\begin{tabular}{@{}lllllll@{}}
\toprule
Experiment & Energy (GeV) & Measurements          & Reported unit & Location               & Altitude        & Zenith range                           \\ \midrule
AMS-02     & 0.1-2500           & Flux \& charge ratio & rigidity      & 28.57$^\circ$N , 80.65$^\circ$ W     & 5 m (sea level) &                                                                          \\
BESS-TeV   & 0.6-400            & Flux                  & momentum      & 36.2$^\circ$N, 140.1$^\circ$W        & 30 m            & 0-25.8$^\circ$                 \\
CMS        & 5-1000             & Charge ratio          & momentum      & 46.31$^\circ$N, 6.071$^\circ$E       & 420 m           & $p\cos\theta_z$                                         \\
L3+C       & 20-3000            & Flux \& charge ratio & momentum      & 46.25$^\circ$N,  6.02$^\circ$E       & 450 m           & 0-58$^\circ$                                                                     \\
MINOS      & 1000-7000          & Charge ratio          & total energy  & 47.82$^\circ$N, 92.24$^\circ$W & 5 m (sea level) & unfolded                           \\
OPERA      & 891-7079           & Charge ratio          & total energy  & 42.42$^\circ$N, 13.51$^\circ$E   & 5 m (sea level) & $E\cos\theta^*$                                 \\ \bottomrule
\end{tabular}
\end{adjustbox}
\caption{List of measurements used for calibration. Most data are taken for vertical incidence angles, or, corrected to vertical through model-dependent unfolding. At this stage, we are using data near sea level. A few more data sets are available from high-altitude balloon flights and near horizontal directions, which we aim to include in the future.}
\label{table:experiments}
\end{table}

We include muon fluxes from L3+cosmic \cite{Achard:2004ws}, Bess-TeV \cite{Haino:2004nq} and AMS-02\footnote{While the recent AMS-02 reference is not a peer-reviewed journal publication, we were curious about the impact of this measurement that comes with small errors, a large energy range and detailed systematic uncertainties} \cite{Duranti:2012css}. Charge ratio measurements come from CMS \cite{Khachatryan:2010mw}, OPERA \cite{Agafonova:2014mzx}, MINOS \cite{Adamson:2007ww}, and also from L3+cosmic and AMS-02. The collection of data covers an energy range from below 1~GeV to 7~TeV and altitudes range from sea level to 450~m. Only L3+C reports measurements at various zenith ranges. The remaining experiments report either at almost vertical angles or have unfolded their angular distribution, reporting a vertical equivalent measurement. We note that this practice, while appealing, introduces an additional model dependence to the data. Reporting the measurement prior to this correction would greatly benefit later analyses.

\section{Data analysis and calibration scheme}
\label{sec:scheme}
We created a dedicated data analysis framework that enables us to accurately include the detailed systematic uncertainties reported by each experiment in conjuncture with the database of individual fluxes and their correction functions for each measurement. In the code, each experiment is implemented as a set of functions for retrieving the published data with errors; correcting the data given a set of systematic functions provided by the experiments (if available); obtaining a flux expectation and comparing it with the corrected data using a $\chi^2$ function. The function includes penalty terms for deviations on the 24 systematic functions considered. The minimization is performed using the code i{\sc Minuit} \cite{iminuit}. The flux database is generated as described in Section~\ref{sec:flux_computation}.

A scheme that adds sufficient degrees of freedom to the flux calculations and captures all hadronic uncertainties is a topic of active discussions. We originally started following the breakdown in $\mathcal{B}_i$ parameters adopted from Barr et al.~\cite{Barr:2006it}. In the initial studies, we attempted to fit the calculated muon fluxes to the combined data using the high-energy parameters (G, H, W and Y), and found their effects to be strongly correlated. Muon flux and charge ratio data are not sufficient to separate the impact of different regions. Particularly problematic are the parameter combinations related to the same projectile energy (such as G and H). We followed multiple steps to simplify the scheme, which resulted in only four parameters that scale the total yield of each charged meson independently ($c_{\pi^+}$, $c_{\pi^-}$, $c_{\text{K}^+}$ and $c_{\text{K}^-}$). This parameterization was found to be more stable against small variations in the input data and, more importantly, is able to describe the data as well as the complicated scheme with twice the number of free parameters.

\section{Experimental considerations}
\begin{figure}
\centering
\includegraphics[width=0.7\textwidth]{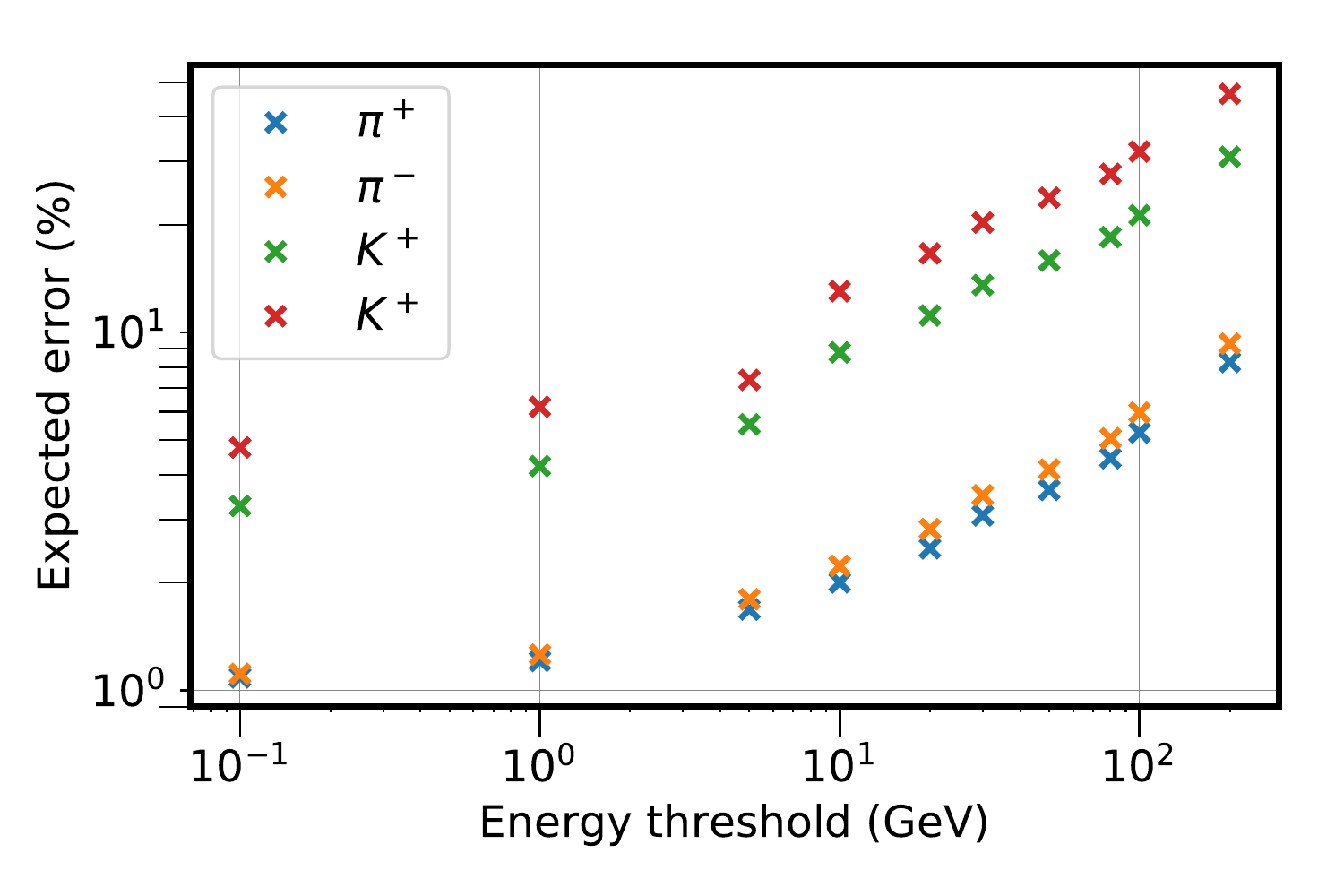} 
\caption{Expected precision of the calibration parameters as a function of the energy threshold. Clearly, the low energy data is very valuable in constraining the contribution of each of the parameters.}
\label{fig:err_vs_e}
\end{figure}

The experiments used cover an energy from a few GeV to several TeV. We find that the low-energy data plays a crucial role in the precise determination of the correction values, as well as in breaking correlations between them. At the same time, we expect the impact of the accurate modeling of the cosmic ray flux, including a geo-magnetic cutoff and solar modulation, to become relevant at very low energies. As demonstrated in \figu{err_vs_e}, a value of 5\,GeV for the energy cutoff below which we discard data is an optimal trade-off where data and calculations can reach an agreement and the modifications introduced are stable.

With the energy threshold fixed to 5\,GeV we also tested the constraints that single experiments could provide on their own. We find that pion yields are mainly constrained by Bess-TeV and L3+cosmic data, while for the kaons, L3+cosmic, MINOS and OPERA are the main contributors to the fit. By removing one experiment at a time from the analysis, we find that L3+cosmic has the largest impact, most likely because it covers regions that no other single experiment can access in energy and arrival direction. 

The same exercise was conducted on the data. We find that the inclusion or exclusion of single experiments can significantly change the overall agreement between the calculations and data. The main contributors to these changes are AMS-02, with rather small errors below 50~GeV and a poor agreement with data above that energy, and CMS, which has data points that significantly deviate from models and experiment below 20~GeV.

\section{Results}
\begin{figure}
\centering
\includegraphics[width=\textwidth]{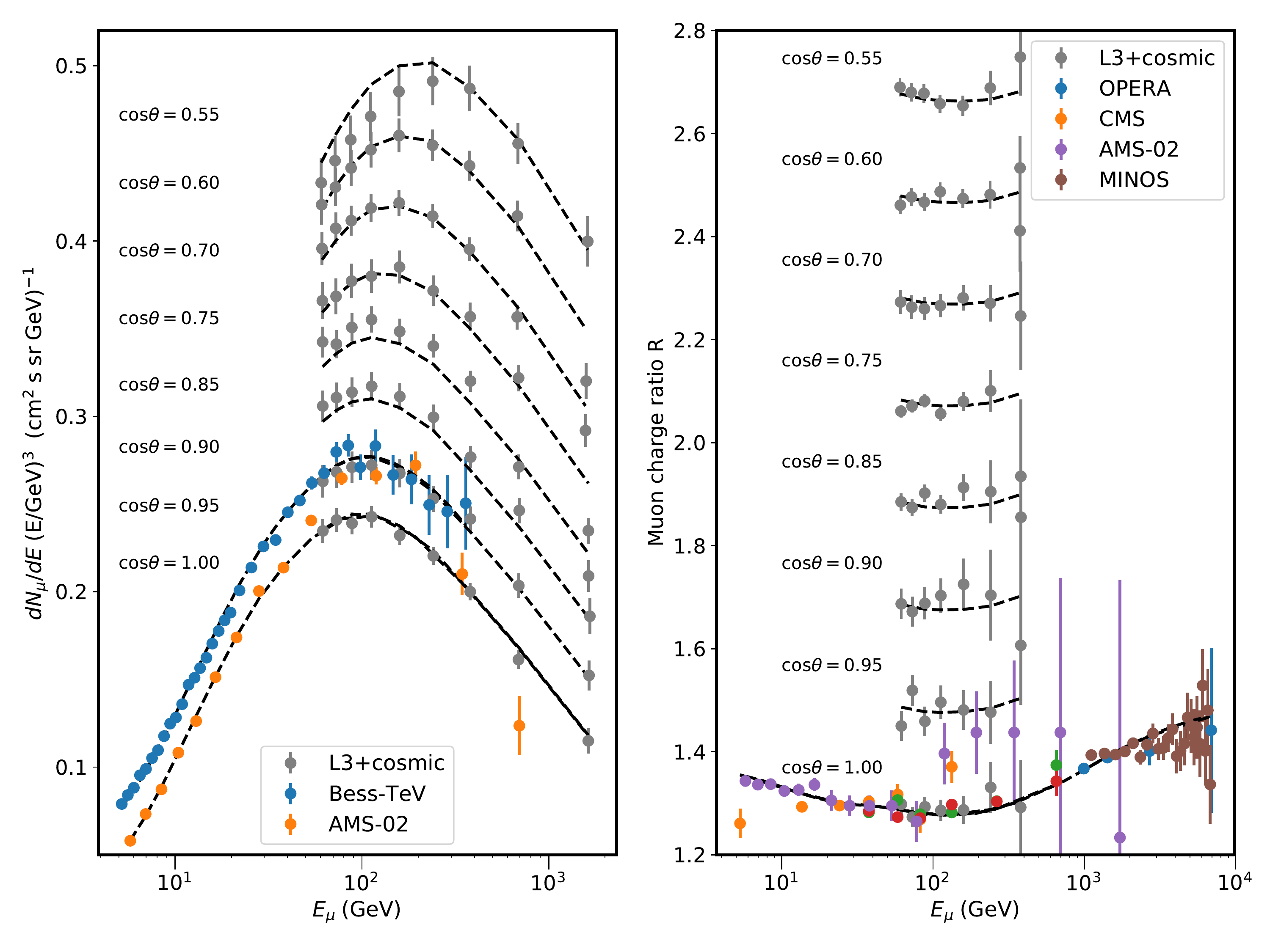} 
\caption{Left panel: Muon flux for the sum of both charges compared to the \sibyll{-2.3c} and GSF flux calculation with corrections from Table \ref{table:results} applied. For the two most vertical zenith angles the data from AMS-02 and Bess-TeV overlap with L3+cosmic. Some occurs between AMS-02 and L3+cosmic above 50 GeV, while for Bess-TeV the agreement is excellent within the errors. There are signs of a systematic disagreement between the calculated and observed angular dependence of the spectrum that is not covered by the systematic uncertainties provided by the experiment. Right panel: Muon charge ratio. As mentioned in the text, most data are unfolded to strictly vertical zenith angles. The agreement across the entire energy range is excellent, except for one of the three CMS measurements. The zenith dependence from L3+cosmic is described well by the corrected model.}
\label{fig:model_vs_exp}
\end{figure}

The result of the fit is demonstrated in \figu{model_vs_exp}. For vertical directions the corrected model matches the data very well. The zenith dependence, presently only available from L3+cosmic data, shows some systematic deviation. We speculate that this may come from a systematic uncertainty not covered by the available correction functions, since none of the performed checks and variations of the fit conditions resulted in a better description of this aspect. The muon charge ratio is also very well described with the exception of one of the three CMS measurements. The charge ratio is the most relevant in constraining the $c_{\text{K}^+}$ parameter.
\begin{figure}
\centering
\includegraphics[width=0.9\textwidth]{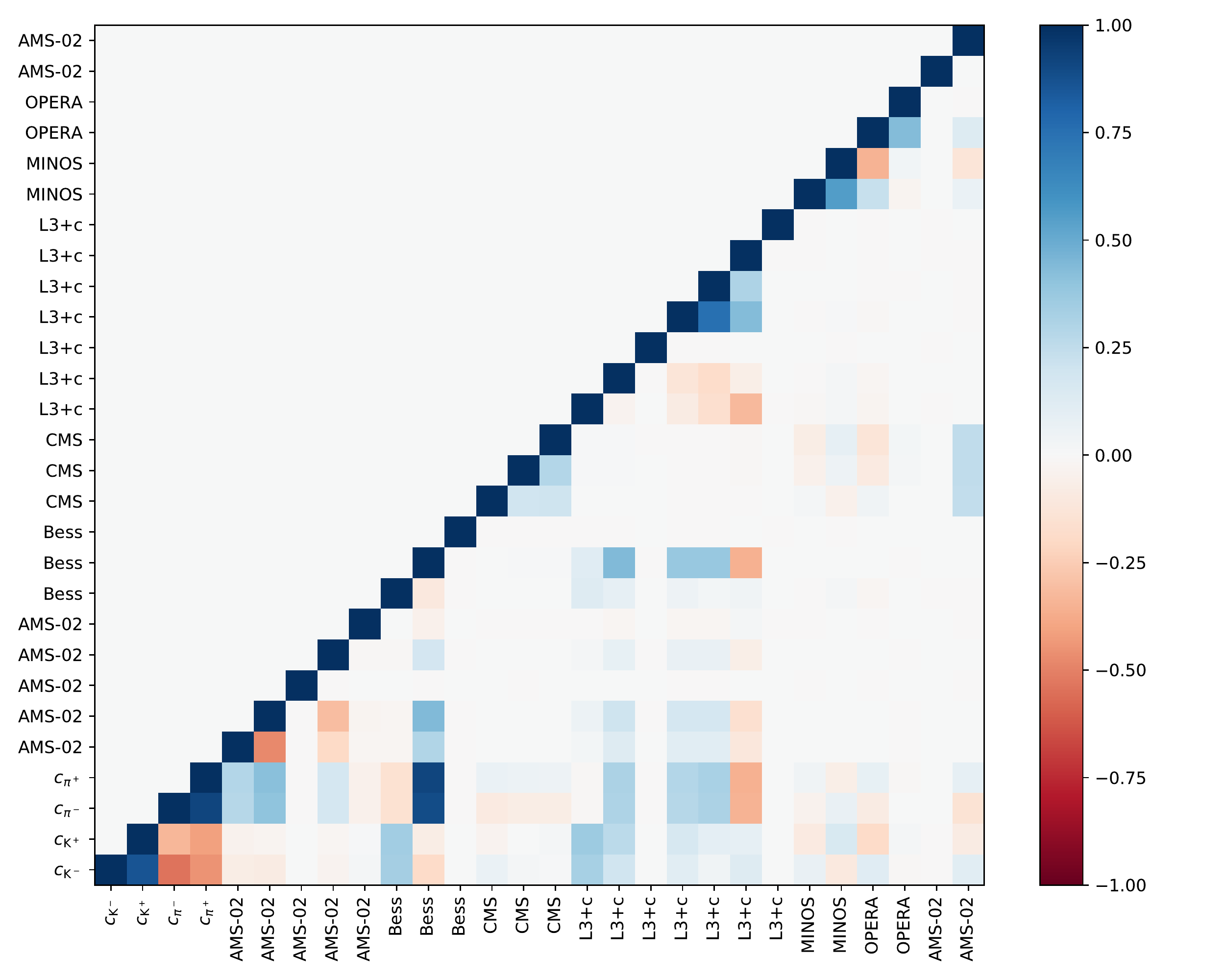} 
\caption{Correlation matrix for the calibration and systematic parameters involved in the fit. Different systematic uncertainties of the same experiment carry the same label for the sake of clarity. As the lower left block demonstrates, the calibration parameters possess stronger correlations, which are smaller compared to schemes with eight and more parameters or fits with a higher energy threshold.}
\label{fig:corr}
\end{figure}

As we argued in Sec.~\ref{sec:scheme}, the fit parameters typically have strong correlations. This is apparent in the lower left block in of the correlation matrix in Fig.~\ref{fig:corr}.
\begin{table}
\centering
\begin{tabular}{@{}lll@{}}
\toprule
Parameter & Best fit & Error      \\ \midrule
$c_{\pi^-}$              & +0.141   & $\pm$0.017 \\
$c_{\pi^+}$              & +0.116   & $\pm$0.016 \\
$c_{\text{K}^-}$               & +0.402    & $\pm$0.073  \\
$c_{\text{K}^+}$               & +0.583    & $\pm$0.055  \\ \bottomrule
\end{tabular}
\caption{Values and errors for the determined correction parameters for the combination \sibyll{-2.3c} and GSF. For \dpmjet{-III-19.1} the pion parameters are similar, but kaon parameters are slightly higher.}
\label{table:results}
\end{table}
The obtained values for the calibration parameters $c_i$ are listed in Table \ref{table:results}. The pion corrections are at the level of 10-15\%. The kaon parameters suggest much higher corrections at the level of 50\%. These values are determined with a precision of about 10\% and 10-20\%, respectively.

\section{Discussion and Outlook}
In this work, we are developing a calibration scheme for high-precision atmospheric neutrino flux calculations. Our aim is to reduce the uncertainties arising from the modeling of hadronic interactions and cosmic ray fluxes by deriving corrections from high-quality atmospheric muon data. 
Initially, it was envisioned to obtain constrains on the sub-divisions of the particle production phase-space according to \cite{Barr:2006it}. However, the many parameters resulted to be strongly correlated and not constrained by world's atmospheric muon data. A drastic simplification to only four parameters resulted in smaller correlations, a more stable fit and a similarly good description of data. 
With this simplified scheme, the derived corrections suggest an increase of roughly 10\% for pions, and an increase of 50\% for kaons. These values correspond to corrections of the particle production in the specific phase-space relevant for atmospheric lepton fluxes ($x_{\rm lab} > 0.1$) and do not necessarily imply an increase of central multiplicities by the same amount. While the errors of these values are small, the current best fit is most likely dominated by systematic effects that are not yet considered, such as uncertainties in the cosmic ray flux. It is also noteworthy that some regions in experimental data cannot be described, regardless of modifications to their correction functions and our model. We will further investigate the impact of these disagreements and reconsider using parts of the data in future. Furthermore, we aim to study the impact of measurements made at high altitude by balloon experiments and data taken at near-horizontal directions.

\bibliographystyle{ICRC}
\bibliography{references}

\end{document}